\documentclass{article}

\usepackage[utf8]{inputenc}
\usepackage[T1]{fontenc}

\usepackage[margin=1.6in]{geometry}

\usepackage{amsthm, amsmath, amssymb}

\usepackage{mathtools}

\usepackage{needspace}
\usepackage{float}
\usepackage{thmtools, thm-restate}
\usepackage{enumerate}
\usepackage[hypertexnames=false]{hyperref}  
\usepackage{graphicx}
\usepackage{authblk} 
\usepackage{titlesec} 

\input{preamble}

\makeatletter
\@ifundefined{newcounteralias}{}{%
  \renewcommand\thmt@autorefsetup{%
    \@xa\def\csname\thmt@envname autorefname\@xa\endcsname
      \@xa{\thmt@thmname}}%
}
\makeatother

\declaretheorem[parent=section, name=Theorem]{theorem}
\declaretheorem[name=Question]{question}

\newtheorem{proposition}[theorem]{Proposition}
\newtheorem{observation}[theorem]{Observation}
\newtheorem{lemma}[theorem]{Lemma}
\newtheorem*{lemma*}{Lemma}
\newtheorem{corollary}[theorem]{Corollary}

\newcommand{\myparagraph}[1]{\paragraph{#1}}

\newcommand{\NP}{{\sf NP}}
\newcommand{\XP}{{\sf XP}}
\newcommand{\FPT}{{\sf FPT}}

\newcommand{\K}{\mathcal{K}}
\renewcommand{\H}{\mathcal{H}}

\renewcommand{\a}{\mathbf{a}}
\newcommand{\GF}{\mathbf{F}_2} 
\newcommand{\one}{\mathbf{1}}  
\newcommand{\zero}{\mathbf{0}} 
\newcommand{\prox}{\,\Tilde{\cap}\,}

\DeclareMathOperator{\GC}{{COMP}}

\newcommand{\red}{\textsf{red}}
\newcommand{\blue}{\textsf{blue}}

\def\FIGmultigraph{0.8}  
\def\FIGsubgraph{0.8}  
\def\FIGbfs{0.8}  
\def\FIGextension{0.8}  

\renewenvironment{abstract}
{\small\vspace{-1em}
\begin{center}
\bfseries\abstractname\vspace{-.5em}\vspace{0pt}
\end{center}
\list{}{
\setlength{\leftmargin}{0.6in}%
\setlength{\rightmargin}{\leftmargin}}%
\item\relax}
{\endlist}

\newif\iflongversion
\longversiontrue
\newcommand{\iflongelse}[2]{\iflongversion{#1}\else{#2}\fi}

\newenvironment{longproof}[1]{\begin{proof}#1}{\end{proof}}

\title{%
On the enumeration of signatures of XOR-CNF's\thanks{%
The authors have been supported by the ANR project PARADUAL (ANR-24-CE48-0610-01). 
An extended abstract of this work has been presented to the WADS 2025 conference~\cite{extendedabstract}.}}

\author{Nadia Creignou}
\author{Oscar Defrain}
\author{Frédéric Olive}
\author{Simon Vilmin}

\affil{Aix-Marseille Université, CNRS, LIS, Marseille, France.}

\date{November 16, 2026} 

\begin{document}

\maketitle

\vspace{-.5cm}

\begin{abstract}
Given a CNF formula $\varphi$ with clauses $C_1, \dots, C_m$ over a set of variables~$V$, a truth assignment $\a\colon V \to \{0, 1\}$ generates a binary sequence $\sigma_\varphi(\a)=(C_1(\a), \ldots, C_m(\a))$, called a signature of $\varphi$, 
where $C_i(\a)=1$ if clause $C_i$ evaluates to 1 under assignment $\a$, and $C_i(\a)=0$ otherwise.
Signatures and their associated generation problems have given rise to new yet promising research questions in algorithmic enumeration.
In a recent paper, Bérczi et al.~proved that enumerating all  signatures of a CNF 
is tractable, despite the fact that verifying that a binary sequence is a signature  is hard.
They also showed the hardness of finding maximal signatures of an arbitrary CNF due to the intractability of satisfiability in general.
Their contribution leaves open the problem of efficiently generating maximal signatures for tractable classes of CNFs, i.e., those for which satisfiability can be solved in polynomial time.
Stepping into that direction, we completely characterize the complexity of generating all, minimal, and maximal signatures for formulas formed by
XOR-clauses, also known as affine formulas from Schaefer’s dichotomy result.

\vskip5pt\noindent{}{\bf Keywords:} algorithmic enumeration, XOR-CNF, signatures, maximal bipartite subgraphs enumeration, extension, proximity search.
\end{abstract}

\section{Introduction}

Propositional formulas are ubiquitous in computer science.
The complexity of the satisfiability problem has been extensively studied from the point of view of several algorithmic tasks such as decidability, counting, or enumeration to mention but a few. 
Given a CNF formula $\varphi$ with clauses $C_1, \dots, C_m$ over a set of variables $V = \{v_1, \dots, v_n\}$, a truth assignment $\a\colon V \to \{0, 1\}$ leads to a binary sequence $\sigma_\varphi(\a)$, called a \emph{signature} of $\varphi$, defined by $\sigma_\varphi(\a)=(C_1(\a), \ldots, C_m(\a))$ where $C_i(\a)=1$ if clause $C_i$ evaluates to 1 under assignment $\a$,  and $C_i(\a) = 0$ otherwise. 
Deciding the satisfiability of $\varphi$ boils down to deciding whether the all-one sequence is a signature of $\varphi$. 

In this paper, we investigate the problems of listing all, minimal, and maximal signatures of a CNF, where minimal and maximal are meant bitwise.
The task of finding all signatures of a CNF originates from well-designed pattern trees and has first been posed during the Dagstuhl seminar on enumeration in data management~\cite{dagstuhl2019db}.
However, enumerating the signatures of a given CNF---may it be all, minimal, or maximal ones---will generally produce an exponential number of solutions. 
Therefore, input sensitive complexity is not a suitable yardstick of efficiency when analyzing enumeration algorithms performance.
Instead, \emph{output-sensitive complexity} estimates the running time of an enumeration algorithm using both input and output sizes.
In that framework, an algorithm is considered tractable when it runs in \emph{total-polynomial time} (or \emph{output-polynomial} time), that is, if its execution time is bounded by a polynomial in the combined size of the input and the output.
Still, further constraining this notion to guarantee some regularity in the enumeration is sometimes desirable.
For this reason, \emph{polynomial delay} and \emph{incremental-polynomial time} are customarily regarded as better notions of tractability for enumeration complexity.
On the one hand, polynomial delay means that the delay between consecutive outputs is polynomial in the input size.
Incremental-polynomial time, on the other hand, means that the time to produce the $i^\text{th}$ solution is bounded by a polynomial in the input size plus $i$ (see e.g.,~\cite{johnson1988generating,strozecki2019survey}).

Generating all the signatures of a CNF turns out to be of great fundamental interest, for it provides cases of enumeration problems that can be solved in total-polynomial time even though solutions cannot be recognized in polynomial time, as we detail now.\footnote{These problems thus lie outside of the class {\sf EnumP} which is commonly studied in algorithmic enumeration; see e.g.~\cite{strozecki2019survey}.}
In a recent contribution~\cite{berczi2021generating}, Bérczi et al.~give an incremental-polynomial time algorithm for listing all the signatures of a $k$-CNF for fixed $k$, while deciding whether the all-one sequence is a signature of a $3$-CNF is hard.
Moreover, they show that for tractable CNF's, signatures can be enumerated with polynomial delay.
On the other hand, they show that generating maximal signatures is hard, while minimal signatures can be listed with polynomial delay, regardless of the CNF under consideration.
Their positive result relies on the equivalence with maximal independent sets enumeration in graphs, known to be tractable~\cite{tsukiyama1977new}. 
As of their hardness result, it relies on the fact that the existence of at least two solutions would imply the non-satisfiability of the CNF at hand.
Thus, the difficulty of the enumeration heavily relies on the intractability of the satisfiability of CNF's in general.
This observation naturally leads to the following open question, where by tractable CNF's we mean families of CNF's for which the satisfiability of a formula $\varphi$ in the family (as well as its sub-formulas) can be decided in polynomial time: \emph{can the maximal signatures of tractable CNF's be enumerated in total-polynomial time?}

Among tractable formulas, Horn and 2-CNF's are two natural candidates for approaching this question.
These two cases however proved to be very challenging, and the question of their tractability was explicitly stated as an open problem in the 2022's edition of the Workshop on Enumeration Problems and Applications (WEPA'22).

Another natural step toward answering the above question is to consider XOR-CNF formulas, being conjunctions of clauses using the  ``exclusive-or'' connective instead of the usual ``or'' connective. 
XOR-clauses were revealed in Schaefer's famous complexity classification as a tractable case for deciding the satisfiability of a set of generalized clauses \cite{Schaefer78}.
Interestingly, XOR-clauses lie at the heart of all  tractable cases when it comes to counting the number of satisfying assignments \cite{CreignouH96}. They subsequently appeared in many Schaefer-like classifications for a variety of algorithmic tasks, including enumeration \cite{CreignouH97, CreignouOS11}.
Many computational problems over XOR-clauses can be solved efficiently due to algebraic properties.
Indeed, a set of such clauses can be seen as a system of linear equations over the 2-element field~$\GF$. As a result, the sets of satisfying assignments of such subsets are affine subspaces, and they allow for algebraic tools such as Gaussian elimination and matroid algorithms. Also in the special case of 2-XOR clauses, subsets of clauses can be interpreted as graphs with 0/1-labeled edges and benefit from several efficient graph algorithms, e.g., for computing connected components or identifying cycles \cite{CreignouOS11}.
This ability to capture XOR clauses using algebraic, matroid, and graph concepts and methods provides efficient tools for analyzing their behavior with respect to our enumeration problem. 

Note that these formulas with unbounded clause length do not fall in the framework of CNF's though.
Indeed, rewriting an XOR-with unbounded clause length CNF as a CNF may result in a blowup on the number of clauses and signatures, as discussed in Section~\ref{sec:preliminaries}.
Hence, we remark that the positive results from~\cite{berczi2021generating} do not directly apply to XOR-CNF.
Yet, we show that signatures enumeration remains tractable using flashlight search as it is done in~\cite{berczi2021generating}. 
Namely, we prove the following.

\begin{restatable}{theorem}{thmxorallsig}
\label{thm:xor-all-signatures}
The set of signatures of an XOR-CNF formula can be enumerated with polynomial delay and polynomial space.
\end{restatable}

Concerning the enumeration of minimal signatures, we show that the problem is in fact equivalent to the case of listing maximal signatures.

\begin{proposition}\label{prop:xor-min-max}
    There is a polynomial-delay and polynomial-space algorithm listing all minimal signatures of an XOR-CNF formula if and only if there is one listing all maximal signatures of an XOR-CNF formula.
\end{proposition}

This leaves us with the study of maximal signatures enumeration for XOR-CNF's.
Quite surprisingly, we prove this problem to relate well with other problems from graph theory and matroid theory.
Specifically, we show the case of 2-XOR-CNF to generalize maximal bipartite subgraphs enumeration, a problem which was recently shown to admit a polynomial-delay algorithm~\cite{conte2022proximity}.
As of the general case of XOR-CNF's, it can be seen as listing all maximal satisfiable subsystems of a linear system of equations, which is related to the enumeration of  circuits passing through a given element in a matroid~\cite{boros2003algorithms}; see also \cite{khachiyan2009generating,khachiyan2005complexity}.
Relying on these results, we derive that maximal signatures enumeration for XOR-CNF's is tractable.
Namely, we obtain the following.

\begin{restatable}{theorem}{thmxormax}
\label{thm:xor-max}
There is an incremental-polynomial time algorithm generating the maximal (or minimal) signatures of an XOR-CNF.
\end{restatable}

Concerning the case of 2-XOR-CNF, we show it can be solved even more efficiently using proximity search, with a polynomial delay.

\begin{restatable}{theorem}{thmtwoxormax}
\label{thm:2-xor-max}
There is a polynomial-delay algorithm generating the maximal (or minimal) signatures of a 2-XOR-CNF.
\end{restatable}

To prove Theorem~\ref{thm:2-xor-max}, we represent the input 2-XOR-CNF as an edge-bicolored graph where each color codes the parity of the XOR clause. 
Then, the enumeration amounts to listing maximal bipartite subgraphs with additional constraints on colored edges.

A caveat of Theorems~\ref{thm:xor-max} and \ref{thm:2-xor-max} is that a queue of solutions has to be maintained in the generation. 
More precisely, in~\cite{boros2003algorithms} the queue is needed to generate a new solution as the algorithm uses the saturation technique where new solutions are derived from obtained ones.
In~\cite{conte2022proximity}, the queue has the other purpose of preventing repetitions.
In both cases, getting rid of a potentially exponential space use is a natural enhancement one may seek.

It should however be noted that improving the algorithm from~\cite{boros2003algorithms} to get polynomial space is open for two decades now.
As of improving Theorem~\ref{thm:2-xor-max} to use polynomial space, it would imply such a result for maximal bipartite subgraphs enumeration, another open question~\cite{conte2022proximity}.
Toward such directions, we prove that the folklore technique of flashlight search used to obtain Theorem~\ref{thm:xor-all-signatures} may probably not be of great use as deciding whether a subsequence of clauses can be extended into a signature is \NP-complete.

\myparagraph{Organization of the paper.} 
\iflongelse{
    In the next section we introduce the notions that will be used in this paper and basic properties.
    The equivalence between maximal and minimal signatures enumeration for XOR-CNF and the proof of Theorem~\ref{thm:xor-all-signatures} are given in Section~\ref{sec:xor-all-signatures}.
    Theorems~\ref{thm:xor-max} and \ref{thm:2-xor-max} are proved in Section~\ref{sec:xor-max-signatures}.
    Future directions and the limitations of flashlight search toward an improvement of Theorem~\ref{thm:2-xor-max} are finally discussed in Section~\ref{sec:discussion}.
}{
    In the next section we introduce the notions that will be used in this paper and basic properties.
    However, due to the space constraint, the proofs of the different statements provided in this extended abstract are omitted.
    The interested reader may still find the arguments in the long version of the manuscript in appendix.
    The steps to prove the equivalence between maximal and minimal signatures enumeration for XOR-CNF and Theorem~\ref{thm:xor-all-signatures} are given in Section~\ref{sec:xor-all-signatures}.
    The paths toward Theorems~\ref{thm:xor-max} and \ref{thm:2-xor-max} are detailed in Section~\ref{sec:xor-max-signatures}.
    Future directions and the limitations of flashlight search toward an improvement of Theorem~\ref{thm:2-xor-max} are finally discussed in Section~\ref{sec:discussion}.
}

\section{Preliminaries}\label{sec:preliminaries}

Let  $V$ be  a finite set, $\mathbf{2}^V$  denotes  its powerset.
Let $m \in \mathbb{N}$ and let $\sigma \in \{0, 1\}^m$ be a binary sequence.
We write $\sigma[i]$ to denote the value of the $i^\text{th}$ element of $\sigma$.
Furthermore it will be convenient to write $\one(\sigma)$ to denote the set of indices $I \subseteq \{1, \dots, m\}$ whose corresponding values in $\sigma$ is 1; $\zero(\sigma)$ is defined analogously.
The \emph{complement} of $\sigma$ is the binary sequence $\bar{\sigma}$ obtained from $\sigma$ by flipping all the bits, i.e., $\bar{\sigma}[i] = 1 + \sigma[i]$ (mod $2$) for $1 \leq i \leq m$.
If $\sigma, \tau \in \{0, 1\}^m$, we write $\sigma \leq \tau$ if $\sigma[i] \leq \tau[i]$ for all $1 \leq i \leq m$.
In other words, $\leq$ corresponds to the bitwise order, and we write $\sigma < \tau$ if $\sigma \leq \tau$ and $\sigma \neq \tau$.

\myparagraph{Enumeration complexity.}
An enumeration algorithm aims at enumerating a set of solutions of some problem, one after the other, with no repetition.
We already have defined relevant time complexities in the introduction.
However, space is also an important consideration in the analysis of enumeration algorithms.
We assume that the solutions we produce are not stored in memory but rather flashed and discarded.
Thus, when measuring space we only consider the space needed by the algorithm in order to conduct the enumeration, in terms of the input size. 
We refer to~\cite{johnson1988generating,strozecki2019survey} for more details on the complexity of enumeration algorithms.

\myparagraph{Graphs.} 
Given an undirected graph $G$, we write $V(G)$ for its set of vertices and $E(G)$ for its set of edges.
For convenience, we will write $uv$ instead of $\{u,v\}$ for edges of $G$.
A graph $G$ is bipartite if there exists a bipartition $V_1, V_2$ of $V(G)$ such that each edge of $E(G)$ has one endpoint in $V_1$ and the other in $V_2$. 
Equivalently, a graph $G$ is bipartite if and only if it does not contain any odd cycles. 

\myparagraph{XOR-formulas.}
We assume basic familiarity with propositional logic.
Let $V = \{x_1,\allowbreak{} \dots, x_n\}$ be a set of Boolean variables.
A \emph{literal} is a variable 
$x_i$ (positive literal) or its negation $\neg x_i$  (negative literal). 
A \emph{clause} $C$ is a disjunction of literals. 
The size $|C|$ of $C$ is its number of literals.
We say that $C$ is a $k$-clause if $|C| \leq k$.
A formula in conjunctive normal form (or a CNF for short) $\varphi$ 
is a conjunction of clauses, i.e., $\varphi = C_1 \land \dots \land C_m$ where $C_i$ is a clause for all $1 \leq i \leq m$.
It will be convenient to denote by $V(\varphi)$ the variables of $\varphi$.
The (total) \emph{size} of $\varphi$ is defined as $\sum_{1 \leq i \leq m} |C_i|$.
An assignment of the variables in $V$ is a mapping $\a\colon V \to \{0, 1\}$.
Given an assignment $\a$, we write $C_i(\a) = 1$ (resp.~$C_i(\a) = 0$) if $C_i$ evaluates to $1$ (resp.~$0$) under assignment $\a$.
The notations $\varphi(\a) = 1$ and $\varphi(\a) = 0$ are defined analogously.
The formula $\varphi$ is \emph{satisfiable} if there exists an assignment $\a$ of $V(\varphi)$ such that $\varphi(\a) = 1$; it is unsatisfiable otherwise. 

An \emph{XOR-clause} is a clause in which the usual connective ``or'' is replaced by the ``exclusive-or'' connective.
It is well-known that any XOR-clause can be represented as a linear equation $x_1 + \dots + x_k = \varepsilon$, $\varepsilon \in \{0, 1\}$ over the two-element field $\GF$.
In this context, $C_i(\a) = 1$ means that the equation $C_i$ is satisfied by $\a$.
In particular, the $2$-XOR-clause $x_1 + x_2 = 0$ (resp.~$x_1 + x_2 = 1$) is satisfied if and only if $x_1 = x_2$ (resp.~$x_1 \neq x_2$). 
An XOR-clause is \emph{odd} (resp.~\emph{even}) if $\varepsilon = 1$ (resp.~$0$).
Given an XOR-clause $C = (x_1 + \dots + x_k = \varepsilon)$, with $\varepsilon \in \{0, 1\}$, we put $\bar{C} = (x_1 + \dots + x_k = 1 - \varepsilon)$ and call $\smash{\bar{C}}$ the \emph{negation} of $C$.
In other words, the bar operator \emph{changes the parity} of~$C$.
An XOR-CNF $\varphi$ is a conjunction of XOR-clauses. 
Equivalently, $\varphi$ can be seen as a system of linear equations over $\GF$.
For an XOR-CNF $\varphi = C_1 \land \dots \land C_m$, we denote by $\bar{\varphi}$ the XOR-CNF $\bar{C}_1 \land \dots \land \bar{C}_m$ and call it the \emph{inverse} of $\varphi$.
We recall that the satisfiability of an XOR-CNF can be tested in polynomial time~\cite[Theorem 2.18]{crama2011boolean}. 

\myparagraph{Signatures.}
The next notations and terminology are borrowed from \cite{berczi2021generating}.
Let $\varphi= C_1 \land \dots \land C_m$ be an XOR-CNF. 
We refer to the introduction for the definition of a signature, and shall say that an assignment $\a$ of $\varphi$ \emph{produces} the signature $\sigma_\varphi(\a)=(C_1(\a), \ldots, C_m(\a))$. 
We will further drop the subscript $\varphi$ from this notation when it is clear from the context. 
Note that in particular, $\varphi$ is satisfiable if and only if $(1, \dots, 1)$ is a signature of $\varphi$.
A signature $\sigma$ of $\varphi$ is \emph{minimal} if it is minimal with respect to $\leq$ among all signatures of~$\varphi$, or in other words, if $\sigma$ is a minimal element of the partial order induced by $\leq$ on the set of all signatures.
Similarly, we call \emph{maximal} a signature which is maximal for $\leq$ among signatures of $\varphi$.

\iflongelse{
    In this paper, we are interested in the following problems.
    
    \begin{problemgen}
      \problemtitle{\textsc Signatures enumeration for XOR-CNF's.}
      \probleminput{an XOR-CNF $\varphi$.}
      \problemquestion{The set of signatures of $\varphi$.}
    \end{problemgen}
    
    \begin{problemgen}
      \problemtitle{\textsc Maximal signatures enumeration for XOR-CNF's.}
      \probleminput{an XOR-CNF $\varphi$.}
      \problemquestion{The set of maximal signatures of $\varphi$.}
    \end{problemgen}
    
    \begin{problemgen}
      \problemtitle{\textsc Minimal signatures enumeration for XOR-CNF's.}
      \probleminput{an XOR-CNF $\varphi$.}
      \problemquestion{The set of minimal signatures of $\varphi$.}
    \end{problemgen}
}
{
In this paper, we are interested in the problem of enumerating all (resp.~minimal, maximal) signatures of a given XOR-CNF.
 }
Observe that in these problems, and by the previous discussions, we may indifferently consider $\varphi$ as a set of XOR clauses or as a set of linear equations over~$\GF$. A subset of simultaneously satisfiable clauses on one side corresponds to a feasible subset of equations on the other side.
In the rest of the paper, we will assume that $\varphi$ is in this latter form.
In particular, when $\varphi$ is seen as a system of linear equations over $\GF$, there is a one-to-one correspondence between maximal signatures of $\varphi$ and maximal feasible subsets of equations.

Another remark is that any XOR-CNF may be rewritten as a CNF, i.e., the ``exclusive-or'' operator may be rewritten using $\land$ and $\lor$.
For example, the XOR-clause $x_1\neq x_2$ can be rewritten as $(x_1\vee x_2)\land (\bar{x_1}\vee \bar{x_2})$.
We however note that we do not have a bijection between the sets of signatures of the two formulas.
Typically, generalizing the above example with clauses $(x_3\vee x_4)\land (\bar{x_3}\vee \bar{x_4})\land (x_5\vee x_6)\land (\bar{x_5}\vee \bar{x_6})\land \dots$ will produce an XOR-CNF with one minimal signature while the equivalent CNF has exponentially many such signatures.
Though this does not rule out the existence of a reduction between (maximal) signatures of XOR-CNF's to (maximal) signatures of CNF's it should be noted that the status of maximal signatures enumeration is still open for 2-CNF's, a point that is discussed in Section~\ref{sec:discussion}.
As of the case of all signatures, it admits a simple and direct proof using flashlight search, as explained in Section~\ref{sec:xor-all-signatures}.

We finally note that the instances of all (resp.~minimal, maximal) signatures enumeration can be assumed to be without duplicate clauses, at an increasing cost which is polynomial per solution. Indeed, if an instance of $\varphi$ contains duplicate clauses, then one of these clauses is evaluated to 1 if and only if each of its copies is also evaluated to 1.
Hence, for each group of duplicate clauses, we can keep only one representative, solve the problem on the reduced instance, and for each obtained solution, add back the duplicates with the same value as their representative in the solution.
Consequently, we will assume without loss of generality that all formulas are without duplicated clauses in the rest of the paper, i.e., that they are defined as pairwise distinct XOR-clauses.

\section{Signatures of XOR-CNF's and their properties} \label{sec:xor-all-signatures}

We prove that testing whether a binary sequence is a signature (resp.~minimal signature, maximal signature) of an XOR-CNF can be done in polynomial time in the size of the CNF at hand, and use it to show that listing all signatures of an XOR-CNF is tractable.
For the rest of the section, let us fix some XOR-CNF $\varphi = C_1 \land\dots \land C_m$ on variable set $V=\{x_1,\dots, x_n\}$.
A preliminary step is the following.

\begin{proposition} \label{prop:complement}
Let $\varphi$ be an XOR-CNF.
A sequence $\sigma\in \{0,1\}^m$ is a signature of $\varphi$ if and only if $\overline{\sigma}$ is a signature of $\bar{\varphi}$.
In particular, $\sigma$ is a minimal (resp.~maximal) signature of $\varphi$ if and only if $\overline{\sigma}$ is a maximal (resp.~minimal) signature of $\bar{\varphi}$.
\end{proposition}

\begin{longproof}
Let $\a$ be an assignment producing the signature $\sigma$ of $\varphi$.
By definition, we have that $C_i(\a) = 1$
and $C_j(\a) = 0$ for each $i,j\in \{1,\dots, m\}$ such that $\sigma[i]=1$ and $\sigma[j] = 0$.
This can be equivalently written as $\smash{\bar{C}_i(\a) = 0}$ if $\overline{\sigma}[i] = 0$ and $\smash{\bar{C}_i(\a) = 1}$ when $\overline{\sigma}[i] = 1$.
This is in turn equivalent to $\a$ producing the signature $\overline{\sigma}$ of $\bar{\varphi}$.
The second part of the statement follows from the fact that for every sequences $\sigma, \tau \in \{0, 1\}^m$ we have that $\sigma \leq \tau$ if and only if $\overline{\tau} \leq \overline{\sigma}$.
\end{longproof}

From Proposition~\ref{prop:complement} we already derive Proposition~\ref{prop:xor-min-max}.
Given two disjoint subsets $A,B \subseteq \{1,\dots, m\}$, we define 
\[ 
\varphi(A,B) \coloneqq  \left(\bigwedge_{i\in A} C_i \right) \land \left(\bigwedge_{j\in B} \bar{C}_j\right)
\]
to be the formula obtained from $\varphi$ by taking the clauses with index in $A$ and the negation of clauses with index in $B$.
We furthermore set this formula to be defined on variable set $V(\varphi)$ even though some variables in $V(\varphi)$ may not appear in any of the clauses of $\varphi(A,B)$---we want this to ensure that assignments of $\varphi(A,B)$ are well-defined for $\varphi$.
Remark that the size of $\varphi(A,B)$ is at most that of $\varphi$, and that it is defined on precisely $m$ clauses whenever $A,B$ is a bipartition of $\{1,\dots, m\}$.
In the next proposition, we show that testing whether a binary sequence is a signature of an XOR-CNF can be reduced to satisfiability testing.

\begin{proposition} \label{prop:signature-sat}
Let $\varphi$ be an XOR-CNF.
A sequence $\sigma\in \{0,1\}^m$ is a signature of $\varphi$ if and only if the formula $\varphi(\one(\sigma), \zero(\sigma))$ is satisfiable.
\end{proposition}

\begin{longproof}
We start with the only if part.
Assume that $\sigma$ is a signature of $\varphi$.
Then there is an assignment $\a$ that produces $\sigma$.
By definition $\a$ satisfies the clauses $C_i$ such that $\sigma[i] = 1$, and it does not satisfy the clauses $C_j$ such that $\sigma[j] = 0$.
This latter case is equivalent to $\a$ satisfying the clauses $\smash{\bar{C}_j}$.
Hence $\a$ is an assignment which satisfies the formula $\varphi(\one(\sigma), \zero(\sigma))$ as desired.

We move to the if part.
Assume that $\varphi(\one(\sigma), \zero(\sigma))$ has a satisfying assignment $\a$ defined over $V$.
Now since $\a$ satisfies the clauses $C_i$ of $\varphi$ such that $\sigma[i] = 1$, and not the clauses $C_j$ of $\varphi$ such that $\sigma[j] = 0$, we conclude that $\sigma$ is indeed a signature of $\varphi$ as required.
\end{longproof}

Note that signatures, seen as sets of indices whose corresponding values in the binary sequence is 1, are neither closed by subset or superset. In other words, there may exist binary  sequences $\sigma$, $ \tau$ and  $\rho$ with  $\one(\sigma)\subset \one(\tau)\subset \one(\rho) $ such that $\sigma,\rho$ are signatures but not $\tau$.
We nevertheless show that testing the minimality (or maximality) of a signature can be done locally in polynomial time.

\begin{proposition}\label{prop:solution-test}
Deciding whether $\sigma$ is a minimal (resp.~maximal) signature of $\varphi$ can be done in polynomial time in the size of $\varphi$.
\end{proposition}

\begin{longproof}
To check that $\sigma$ is a signature, we rely on Proposition \ref{prop:signature-sat}: we build the XOR-CNF $\varphi(\one(\sigma), \zero(\sigma))$ and test its satisfiability, all of which in polynomial time.
In the rest of the proof, for any $j\in \zero(\sigma)$, let us denote $\psi(\sigma, j)\coloneqq \varphi(\one(\sigma), \emptyset)\wedge C_j$, i.e., $\psi(\sigma, j)$ is obtained as the conjunction of the clauses to $1$ in $\sigma$ plus $C_j$.
We show that $\sigma$ is not a maximal signature of  $\varphi$ if and only if $\psi(\sigma, j)$ is satisfiable for some $j\in \zero(\sigma)$.

We prove the if part. 
Suppose that $\psi(\sigma, j)$ is satisfiable for some $j\in \zero(\sigma)$, and consider an arbitrary assignment $\a$ over $V$ satisfying $\psi(\sigma, j)$. 
Let $\tau$ be the signature of $\varphi$ produced by assignment $\a$.
Since $\a$ satisfies $\psi(\sigma, j)$, every $1$ in $\sigma$ is a $1$ in $\tau$, so that $\sigma \leq \tau$.
Moreover, $\sigma[j]<\tau[j]$ holds, and $\sigma < \tau$ thus follows.
Hence $\sigma$ is not a maximal signature concluding the if implication.
As for the only if part, the existence of a signature $\tau > \sigma$ implies by definition that $\psi(\sigma, j)$ is satisfiable for some $j\in \zero(\sigma)$ such that $\tau[j]>\sigma[j]$.

Hence, deciding whether $\sigma$ is a maximal signature of $\varphi$ can be done by $m$ calls to the satisfiability of an XOR-CNF whose size is no greater than $\varphi$.
We conclude that it can be done in polynomial time.
For minimality, we conclude using Proposition~\ref{prop:complement}.
\end{longproof}

To conclude this section, we prove that enumerating all the signatures of an XOR-CNF is tractable using the above properties and flashlight search.
The main idea of this technique is to conduct binary search over the space of solutions by recursively forcing and excluding vertices.
In order for flashlight search to run with polynomial delay and polynomial space, it is sufficient to solve the so-called ``extension problem'' in polynomial time: given two disjoint subsets $A, B$ of the ground set, decide whether there exists a solution including $A$ and avoiding $B$.
We refer to e.g.~\cite{boros2004algorithms,strozecki2019efficient} for a more detailed description of this folklore technique.
In the case of XOR-CNF signatures, $A$ will consist of indices of clauses to be satisfied in the signature---the 1's in the sequence---while $B$ will consist of indices of clauses not to be satisfied---the 0's in the sequence.

\thmxorallsig*

\begin{proof}
From the above discussion, we have that $\varphi$ admits a signature $\sigma$ satisfying $\sigma[i]=1$ and $\sigma[j]=0$ for all $i\in A$ and $j\in B$ if and only if $\varphi(A,B)$ is satisfiable.
Indeed, a satisfying assignment $\a$ of $\varphi(A, B)$ satisfies the clauses of $\varphi$ with index in $A$, and not those with index in $B$.
Thus the signature of $\varphi$ produced by $\a$ meets the requirements.
As for the converse direction, it follows directly from the definition of a signature.
We conclude by Proposition~\ref{prop:solution-test} that the extension problem can be solved in polynomial time for XOR-CNF signatures, which concludes the proof.
\end{proof}

We point out that Theorem~\ref{thm:xor-all-signatures} is not a direct consequence of~\cite[Theorem 1]{berczi2021generating}. 
Indeed, in the latter theorem the authors consider families of tractable CNF's which do not contain XOR-CNF's in general.
The arguments, however, follow the same line.

\section{Maximal signatures of XOR-CNF's}\label{sec:xor-max-signatures}

We investigate the complexity of enumerating the maximal signatures of a given XOR-CNF. More precisely, we prove the general case to be solvable in incremental-polynomial time, 
and the 2-XOR-CNF case to be solvable with polynomial delay using proximity search.

In~\cite{boros2003algorithms}, the authors present an incremental polynomial-time algorithm for enumerating all circuits of a matroid containing an fixed element. 
We can derive an algorithm that enumerates the maximal signatures of a given XOR-CNF within the same time.

\iflongelse{
    \thmxormax*
}
{
    This immediately yields Theorem~\ref{thm:xor-max}.
}

\begin{proof}
We mentioned in Section~\ref{sec:preliminaries} that each clause of a XOR-CNF $\varphi$ can be seen as a linear equation $x_1 + \dots + x_k = \varepsilon$ over the two-element field~$\GF$. Hence $\varphi$ can be seen as a system $\cal S$ of such equations and maximal feasible subsystems of $\cal S$ correspond to maximal signatures of $\varphi$.
If $\cal S$ is feasible, then there is only one maximal signature. Now, assume $\varphi$ gives rise to an unfeasible system: 
\begin{equation}\label{eq:Ax=b}
Ax = b
\end{equation}
where 
$A\in{\cal M}_{pn}(\GF)$, $b \in{\cal M}_{p1}(\GF)$ and $x$ is a vector of $n$ variables, for some $p,n\in\mathbb N$. 
Now, enumeration of maximal feasible subsystems of~\eqref{eq:Ax=b} can be reduced to enumeration of solutions with minimal support of a suitably transformed system of equations. This is done as follows. Let $y$ be a column vector consisting of $p$ new variables and consider the system: 
\begin{equation}\label{eq:Ax+y=b}
Ax + y = b
\end{equation}
Every maximal feasible subset of equations of system~\eqref{eq:Ax=b} corresponds to the set of $0$-coordinates in a vector $y$ with a minimal support that can be extended to a solution of system~\eqref{eq:Ax+y=b}. This later system can be equivalently written: 
\begin{equation*}
\left[\begin{array}{c|c}
A & I_p
\end{array}\right]
\begin{pmatrix} x \\ y \end{pmatrix} 
= b
\end{equation*}
where $I_p$ is the identity matrix of dimension $p$. Then, in using Gaussian elimination, we get a new equivalent system of the form: 
\newcommand{\Bu}{B}\newcommand{\Bd}{C}
\newcommand{\cu}{u}\newcommand{\cd}{v}
\newcommand{\zM}{\textbf{0}} 
\begin{equation*}
\left[ \begin{array}{c|c} T & \Bu \\ \hline \zM & \Bd \end{array} \right]
\begin{pmatrix} x\\ y \end{pmatrix} 
= 
\begin{pmatrix} \cu \\ \cd  \end{pmatrix}
\end{equation*}
where the matrices above have the following shapes, for some integer $k$: 
\[
T\in{\cal M}_{kn}(\GF),         \ 
\Bu\in{\cal M}_{kp}(\GF),       \ 
\Bd\in{\cal M}_{(p-k)p}(\GF),   \ 
\cu\in{\cal M}_{k1}(\GF),       \ 
\cd\in{\cal M}_{(p-k) 1}(\GF), 
\]
and where furthermore, $T$ is in reduced row-echelon form. 
This last system is the conjunction of the two subsystems~\eqref{eq:u} and~\eqref{eq:v} below: 
\begin{align}
Tx + \Bu y      & = \cu         \label{eq:u} \\
\Bd y           & = \cd         \label{eq:v} 
\end{align}
Observe that, since $T$ is in reduced row-echelon form, every solution $y$ of~\eqref{eq:v} can be extended to a solution $(x,y)$ of the full system~(\ref{eq:u},\ref{eq:v}), and hence, of the system~\eqref{eq:Ax+y=b} introduced earlier. 
Therefore, every  maximal feasible subsystem of our initial system~\eqref{eq:Ax=b} corresponds to a vector $y$ with minimal support that is a solution to the system~\eqref{eq:v}, or, equivalently, that satisfies the equality
\(
y_1\Bd_1+\dots+y_p\Bd_p+\cd  = 0 
\),
where the $\Bd_i$'s are the $p$ column vectors of~$\Bd$.
Finally, listing all $(y_1,\dots,y_p)$ of this kind is equivalent to  listing the circuits of the vectorial matroid $(\Bd_1,\dots,\Bd_p, \cd)$ passing through the element $v$. This task can be completed in incremental-polynomial time, as proven in~\cite{boros2003algorithms}. 
\end{proof}

We now focus on 2-XOR-CNF's. 
Recall from Section~\ref{sec:preliminaries} that we can consider these formulas as a set of 2-clauses of the form $x = y$ or $x\neq y$ for positive literals $x, y$ only, together with a set of 1-clauses $(z)$ with $z$ a positive or negative literal. Furthermore, we may assume that no clause is repeated in the formula, as far as signatures enumeration is concerned. We shall call \emph{graphic} a 2-XOR-CNF with clauses of the first type only---that is, without 1-clauses---and will assume for now on that our XOR-CNF's are graphic.
Later, we will show how to deal with 1-clauses with small modifications on the instance.

We start by defining an underlying colored graph structure of graphic 2-XOR-CNF that will be convenient for applying the framework of proximity search.
In the following, given a graphic 2-XOR-CNF $\varphi$, we call \emph{colored edge} a pair $(\{x,y\}, c)$ where $x,y\in V(\varphi)$, $x\neq y$ and $c\in \{\red,\blue\}$.
We define $G(\varphi)$ (or simply $G$ when it is clear from the context) as the edge-bicolored graph defined on vertex set $V(\varphi)$ and where there is an edge $(\{x,y\}, \blue)$ whenever $x\ne y$ is a clause of $\varphi$, and an edge $(\{x,y\}, \red)$ whenever $x=y$ is a clause of $\varphi$.
Note that we allow both a red edge and a blue edge between a pair of variables $x, y$: this corresponds to having clauses $x=y$ and $x\neq y$ in $\varphi$ which means that a satisfiable subformula contains at most one of these clauses.
However as no clause is repeated in $\varphi$, no two edges of the same color share the same endpoints. 
See Figure~\ref{fig:multigraph} for an example of such a graph.
In the following, we will label colored edges of $G$ as $e_1,\dots,e_m$ so that the edge $e_j$ corresponds to the clause $C_j$ in~$\varphi$.
Moreover, when no two colored edges share the same endpoints (and hence only differ by their color) or when there is no ambiguity, we may identify an edge with its vertices and use classical graph notations for edges, that is use $uv$ instead of $(\{u, v\}, c)$.
We denote by $R(G)$ the set of red edges of $G$, and $B(G)$ the set of its blue edges.
Observe that by definition, $R(G)$ and $B(G)$ define a bipartition of the colored edges of $G$.

In the rest of the section, by \emph{subgraph} of an edge-bicolored graph $G$ we mean a graph $H$ with $E(H)\subseteq E(G)$ and $V(H)\subseteq V(G)$ where $V(H)$ contains at least the vertices involved in the edges of $E(H)$. 
For convenience, we may as well refer to a subgraph by its edge set, in which case we assume its vertex set to be the one induced by the endpoints of its edges.
Finally, given two disjoint subsets $X,Y$ of a graph $G$, by $\delta_G(X,Y)$ we mean the set of edges having an endpoint in $X$ and the other in $Y$, and may drop $G$ from the subscript if no ambiguity arises.

\begin{figure}[ht!]
    \centering
    \includegraphics[page=1, scale=\FIGmultigraph]{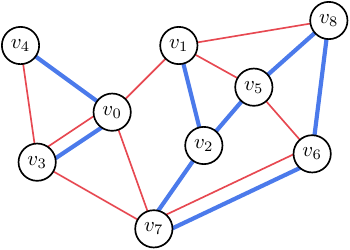}
    \caption{An edge-bicolored graph $G$ on nine vertices, with red edges denoted by thin red lines and blue edges denoted by bold blue lines. For example, the blue edge $v_0v_3$ stands for the clause $v_0\neq v_3$, while the red edge $v_0v_3$ stands for $v_0=v_3$.}
    \label{fig:multigraph}
\end{figure}

Note that an arbitrary (simple, uncolored) graph $H$ is bipartite whenever it admits a bipartition $(X,Y)$ of its vertex set with $E(H)=\delta(X,Y)$.
In the next lemma, we relate the satisfiability of a graphic 2-XOR-CNF to a property of its underlying edge-bicolored graph which generalizes bipartiteness by treating the red and blue edges
differently: blue edges should connect vertices from different parts of a bipartition, while
red edges connect edges in the same part.
More formally, we say that an edge-bicolored graph $G$ is \emph{red-blue bipartitionable}\footnote{In a previous version of this manuscript, this notion was called ``red-blue bipartite.'' Following the suggestion of an anonymous reviewer, we changed the name to avoid any confusion with the notion of bipartite graphs, since red-blue bipartitionable graphs may not need to be bipartite.} if there is a bipartition $(X,Y)$ of its vertex set such that $B(G)=\delta(X,Y)$ or, equivalently, $R(G)=E(G)\setminus \delta(X,Y)$.
In Figure \ref{fig:subgraph}, we give an example of a maximal red-blue bipartitionable subgraph $H$ of the graph $G$ of Figure \ref{fig:multigraph}.

\needspace{1cm}

\begin{lemma}\label{lem:2-xor-sat}
    Let $\varphi$ be a graphic 2-XOR-CNF.
    Then $\varphi$ is satisfiable if and only if $G(\varphi)$ is red-blue bipartitionable.
    Moreover, every bipartition $(X,Y)$ witnessing this property corresponds to a model $\a$ of $\varphi$ with $X=\a^{-1}(1)$ and $Y=\a^{-1}(0)$. 
\end{lemma}

\begin{longproof}
    
    Let $G:=G(\varphi)$.
    Clearly, an assignment of variables $\a$ is a satisfying assignment of $\varphi$ if and only if the sets $X=\a^{-1}(0)$ and $Y= \a^{-1}(1)$ satisfy $B(G) = \delta_G(X,Y)$.
    Moreover, if the bipartitions $(X,Y)$ are considered as ordered pairs and empty parts are allowed, then the relations above define a bijection between bipartitions and assignments of the variables.
\end{longproof}

We derive the following corollary for signatures, where by maximal subgraph we refer to edge inclusion.

\begin{corollary}\label{cor:2-xor-sig}
    Let $\varphi$ be a graphic 2-XOR-CNF and $\sigma\in \{0,1\}^m$.
    Then $\sigma$ is a maximal signature of $\varphi$ if and only if the subgraph $H=\{e_j : e_j\in E(G),\ \sigma[j]=1\}$ of $G(\varphi)$ is a maximal subgraph of $G$ admitting a bipartition $(X,Y)$ such that $\delta_H(X,Y) = B(H)$.
\end{corollary}

\begin{figure}[ht!]
    \centering
    \includegraphics[scale=\FIGsubgraph, page=2]{figures/parcours.pdf}
    \caption{A maximal (connected) red-blue bipartitionable subgraph $H$ of the graph $G$ defined in Figure \ref{fig:multigraph}.
    The bipartition $(H_0, H_1)$ of $H$ is highlighted.
    The edges of $G$ not in $H$ are dotted.}
    \label{fig:subgraph}
\end{figure}

\begin{theorem}\label{thm:2-xor-bip}
    There is a polynomial-delay and polynomial-space algorithm enumerating the maximal signatures of a graphic 2-XOR-CNF $\varphi$ if and only if there is one listing the maximal red-blue bipartitionable subgraphs of its underlying graph $G(\varphi)$.
\end{theorem}

\begin{longproof}
    This is a consequence of Corollary~\ref{cor:2-xor-sig} observing that constructing either the formula or the edge-bicolored graph can be done in polynomial time in the size of the structure at hand.
    As there is a bijection between the two solutions sets this reduction preserves polynomial delay and polynomial space.
\end{longproof}

A consequence of Theorem~\ref{thm:2-xor-bip} is that the enumeration of maximal signatures of a 2-XOR-CNF only consisting of disequalities
$x\neq y$ is equivalent to the enumeration of maximal (edge) bipartite subgraphs of a given graph.
Indeed in that case, $R(G)$ is empty---hence $G$ is a graph---and we only require $E(G)=\delta(X,Y)$.
We derive the following using a recent result by Conte, Grossi, Marino,  Uno and Versari~\cite{conte2022proximity} on maximal bipartite subgraphs enumeration by proximity search.

\begin{corollary}\label{thm:2-xor-bluebip}
    There is a polynomial-delay algorithm listing the maximal signatures of a 2-XOR-CNF consisting of only disequalities $x\ne y$.
\end{corollary}

We conclude the section by showing that the result from~\cite{conte2022proximity} can be extended into our context using proximity search, and by arguing how to capture clauses of size one.

In the following, we say that an edge-bicolored graph $G$ is \emph{connected} if its underlying graph (where colors are ignored and multi-edges are considered as a single edge) is connected.
Note that we can assume without loss of generality that $G(\varphi)$ is connected, as far as signatures enumeration is concerned. 
Indeed, if it was not the case we could enumerate the signature of each subformula of $\varphi$ corresponding to connected components of $G(\varphi)$ and combine the results by doing their Cartesian product.
Hence, we may further assume that $G(\varphi)$ is connected, and shall call \emph{connected} a formula such that its underlying graph is.
We first show that if $G$ is connected then its maximal red-blue bipartitionable subgraphs are as well, yielding a canonical bipartition.

\begin{lemma}\label{lem:red-blue-connected}
    Let $G$ be a connected edge-bicolored graph. 
    Then any of its maximal red-blue bipartitionable subgraphs is connected and spans all vertices.
\end{lemma}

\begin{longproof}
    Let $H$ be an arbitrary maximal red-blue bipartitionable subgraph of $G$ and suppose toward a contradiction that it is not connected or does not span all vertices.
    If $H$ is not spanning, it is a proper subgraph of $H' := (V(G), E(H))$ which is itself a red-blue bipartitionable subgraph of $G$ since isolated vertices can be put in any of the parts defined by $E(H)$.
    Let us then assumed $H$ is not connected.
    Since $G$ is connected we can find a path $P$ connecting two distinct connected components $C_1,C_2$ of $H$. 
    Consider the first edge $e$ not in $H$ on this path from $C_1$ to $C_2$.
    We obtain a red-blue bipartitionable supergraph of $H$ by adding $e$, either keeping its other endpoint in the same part if $e$ is red, or put its endpoint in the other part if it is blue.
    Note that we may need to swap the original bipartition of $C_2$ if that edge connects $C_1$ to $C_2$.
    No contradiction can be found during this process and the obtained graph is indeed a supergraph of $H$.
    This contradicts the fact that $H$ is maximal and we conclude that $H$ is connected and spans all vertices as desired.
\end{longproof}

The next lemma is essentially proved by considering a traverse of the graph initiated on its vertex of smallest label. 

\begin{lemma}\label{lem:red-blue-unique-bipartition}
    If $G$ is a connected edge-bicolored red-blue bipartitionable graph then it admits a unique bipartition $(X,Y)$ of its vertex set with $B(G)=\delta(X,Y)$ and $X$ containing the vertex of smallest label in $G$.
\end{lemma}

\begin{longproof}
    Note that if the assignment of a vertex $u$ into $X$ or $Y$ is determined, and $u$ is contained in a colored edge $(\{u,v\},c)$, then also the assignment of $v$ into $X$ or $Y$ is determined depending on whether the color $c$ is red or blue. Since the graph is connected, determining the assignment of any of its vertices into $X$ or $Y$ determines the assignment of all other vertices.
\end{longproof}

Proximity search has been introduced in~\cite{conte2019proximity} as a special case of solution graph traversal in which the proof that the solution graph is strongly connected relies on a notion of proximity between solutions that is more involved than their intersection. 
More precisely, the proximity between two solutions $S,S^*$ is defined as the length of the longest prefix of $S^*$ (according to a carefully chosen ordering of its elements) that is subset of $S$, and needs not to be symmetric. 
We refer to~\cite{conte2022proximity} for more details on the technique.
In~\cite{conte2022proximity}, the authors extract the necessary conditions that make an enumeration problem ``proximity searchable'' and thus to be solvable with polynomial delay.
We restate these conditions here in the context of XOR-CNF for self-containment and better readability.

In the reformulation below, let $G$ be an edge-bicolored graph and $\H(G)$ denote the set of maximal (connected) red-blue bipartitionable subgraphs of $G$ we shall call \emph{maximal solutions}.
We further call \emph{solutions} the connected red-blue bipartitionable subgraphs of $G$.

\begin{theorem}[Reformulation of {\cite[Definition 4.2 and Theorem 4.3]{conte2022proximity}}]\label{thm:xor-proximity}
    For each $H \in \H(G)$, let $\mu(H)$ denote an ordering of the edges in $H$ and, for arbitrary $H,H^*\in \H(G)$, let $H\prox H^*$ denote the elements in the longest prefix of $\mu(H^*)$ that is a subset of $H$.
    Then, there is a polynomial-delay algorithm enumerating $\H(G)$ whenever:
    
    \begin{enumerate}
        \item For any $H\in \H(G)$, any prefix of $\mu(H)$ is a solution;
        \item There is a polynomial-time computable function $\GC$, which given a solution $H'$ produces   $H=\GC(H')$   a maximal solution such that $H\supseteq H'$; and
        
        \item Given $H\in \H(G)$ and $e\in E(G)\setminus E(H)$ there is a family $\K\subseteq 2^{E(H)}$ of sets called removables such that: 
        \begin{itemize}
            \item the family $\K$ can be computed in polynomial time;
            \item for any $K\in \K$, $H\setminus K \cup \{e\}$ is a solution; and
            \item for any $H^*\in \H(G)$, if $e$ is the minimal element of $\mu(H^*)$ not in $H$, then there exists $K\in \K$ with $(H\prox H^*) \cap K=\emptyset$.
        \end{itemize}
    \end{enumerate} 
\end{theorem}

Let us briefly explain the role of the conditions of Theorem~\ref{thm:xor-proximity}.
Each maximal solution has a polynomial number of neighbors determined by the removables of Item 3 and extended in polynomial time to new maximal solutions using Items 1 and 2.
This defines a solution-graph.
The fact that this solution-graph is strongly connected follows from the fact that for any pair of maximal solutions $H,H^*$, there exists a neighbor of $H$ that increases the proximity with $H^*$.

The rest of the section is dedicated to proving that the conditions of Theorem~\ref{thm:xor-proximity} are fulfilled for well-chosen $\mu$ and set $\K$.
The fact that $\GC$ can be computed in polynomial time follows from the fact that the property of being red-blue bipartitionable is a hereditary property which can be tested in polynomial time, hence that we can greedily obtain a maximal solution. 

We start with the definition of $\mu$.
Given a maximal red-blue bipartitionable graph $H$ of $G$ we denote by $(H_0, H_1)$ the unique bipartition given by Lemma~\ref{lem:red-blue-unique-bipartition} where $H_0$ is the side containing the vertex $v_0$ of smallest index in $G$.
Consider a BFS ordering of the vertices of $G$ initiated at $v_0$: we start with $v_0$ which is unmarked, and at each step, we iteratively consider unmarked vertices ordered so far, and, in order, we mark them and append the unmarked vertices of their neighborhoods in ascending order of their label.
Clearly, each prefix of such an ordering defines a connected (vertex) induced subgraph.
We define $\mu(H)$ as the increasing ordering of the edges of $H$ with respect to their endpoint occurring later in the BFS ordering launched at~$v_0$, and break the tie (two edges sharing the same later endpoint) by increasing order of their earlier endpoint
in the BFS; see Figure~\ref{fig:BFS} for an example of such an edge-ordering.
This choice of an edge-ordering of a maximal solution yields the following property which will be crucial in the rest of the proof, and from which, intuitively, we will obtain that pairs of solutions $H,H^*$ must agree on their respective bipartition of the elements in $H\prox H^*$.

\begin{figure}[ht!]
    \centering
    \includegraphics[scale=\FIGbfs, page=3]{figures/parcours.pdf}
    \caption{A BFS on the maximal red-blue bipartitionable subgraph $H$ of $G$ from Figure \ref{fig:subgraph} will give the following order on the vertices: $v_0, v_1, v_3, v_4, v_2, v_5, v_8, v_7, v_6$.
    The order of the edges of $H$ it produces is $\mu(H) = v_0 v_1, v_0 v_3,\allowbreak{} v_0 v_4,\allowbreak{} v_3 v_4,\allowbreak{} v_1 v_2,\allowbreak{} v_1 v_5,\allowbreak{} v_2 v_5,\allowbreak{} v_1 v_8,\allowbreak{} v_3 v_7,\allowbreak{} v_8 v_6, v_7 v_6$.}
    \label{fig:BFS}
\end{figure}

\begin{observation}\label{obs:prefix}
    If $H\in \H(G)$ then any prefix of $\mu(H)$ is a connected red-blue bipartitionable subgraph of $G$ and the first edge in this ordering is incident to $v_0$.
\end{observation}

Let us now consider arbitrary $H,H^*\in \H(G)$ and put $H'\coloneqq H\prox H^*$ as defined in Theorem~\ref{thm:xor-proximity}. 
Note that if $H'$ is empty then the trivial removable $K=E(H)$ satisfies the above conditions.
Hence we may assume for convenience (in the rest of the analysis) that $|H\prox H^*| \neq 0$, while it will appear clear later that this assumption is in fact not needed.
By Observation~\ref{obs:prefix}, since the first edge in $\mu(H^*)$ belongs to both $H'$, $H$ and $H^*$, we derive that $v_0$ belongs to all three graphs.
Again by Observation~\ref{obs:prefix}, $H'$ is a connected red-blue bipartitionable subgraph of $G$.
Thus by Lemma~\ref{lem:red-blue-unique-bipartition} it admits a unique bipartition $(H'_0, H'_1)$
with blue edges lying between $H'_0$ and $H'_1$,
and such that $v_0\in H'_0$.
Now, since $H$ and $H^*$ contain $H'$, they must agree with that bipartition, meaning that $H'_0\subseteq H_0\cap H^*_0$ and $H'_1\subseteq H_1\cap H^*_1$, where $(H_0^*, H_1^*)$ denote the unique bipartition of $H^*$ such that $v_0 \in H_0^*$ 
as provided by Lemma~\ref{lem:red-blue-unique-bipartition}.
In the following, for each edge $e = ab$ in $E(G) \setminus E(H)$, we use its edges incident to the vertices of $e$ to build the removables.
More precisely, we define $K_1\coloneqq \{av : av\in E(H)\}$ and $K_2\coloneqq \{bv : bv\in E(H)\}$ to be the two removables of $(H,e)$
and argue that $\K\coloneqq \{K_1,K_2\}$ meets the requirements of the theorem.
We stress the fact that in $K_1$ or $K_2$ we may find edges lying in $H^*$, which is not a problem; rather, we will argue that, when choosing $e$ as the minimal element of $\mu(H^*)$ not in $H$, one of $K_1$ or $K_2$ will be disjoint from $H\prox H^*$, hence that $\K$ defines a valid set of removables.

Consider the first edge $e=ab$ in $\mu(H^*)$ that is not an edge of $H$. 
By construction, every edge in $\mu(H^*)$ except of the first one has at least one vertex
in common with some of the previous edges. If $H'$ is not empty, this implies that
at least one of the vertices $a,b$ belongs to $H'$. Let us first prove that the
other vertex is not in $H'$. 
Assume for a contradiction that both vertices are in $H'$.
Suppose first that $ab$ is a blue edge. 
Since $H$ is maximal and does not contain $ab$,
the vertices $a,b$ lie in the same part in $H$ and, hence, also in $H'$. 
Since $H^*$ contains $ab$, the vertices $a,b$ lie in different parts of $H^*$ and, hence, also in $H'$.
This is a contradiction, so one of the vertices $a,b$ is not a vertex of $H'$. 
The same arguments yield the contradiction if $ab$ is a red edge. 
As a consequence, we obtain that one of the vertices $a,b$, namely the one not in $H'$, satisfies that none of the edges incident to it is an edge of $H'$. Hence, one of the $K_0$ and $K_1$ is disjoint with~$H'$.

We are ready to conclude with Theorem~\ref{thm:2-xor-max} where we prove that a $2$-XOR-CNF with singleton clauses can be reduced to the case of graphic XOR-CNFs.

\iflongelse{
    \thmtwoxormax*
}
{
    We conclude with Theorem~\ref{thm:2-xor-max} as a consequence of Theorem~\ref{thm:xor-proximity} and the above discussion.
}

\begin{proof}
By Theorems~\ref{thm:2-xor-bip}, \ref{thm:xor-proximity} and above discussions, it remains only to argue that we can reduce the problem of enumerating the maximal signature of a $2$-XOR-CNF with singleton clauses to the problem of finding the maximal red-blue bipartitionable subgraphs of some edge-bicolored graph $G$.
Let $\varphi$ be such a $2$-XOR-CNF.
As a preliminary step, note that 1-clauses containing variables that do not appear in 2-clauses can be removed from the instance as they will be set to true in any maximal signature.
Thus we may assume that $\varphi$ does not contain such isolated 1-clauses.
We code every other 1-clauses $(x)$ and $(\overline{y})$---which can also be seen as clauses $x=1$ and $y=0$---by a blue edge $xu$ and a red edge $yu$ in the edge-bicolored graph $G$ we have defined above, for some special additional vertex $u$ that will be connecting all such 1-clauses. 
Note that $G$ stays connected by the above assumption that $\varphi$ has no isolated 1-clause.
Then we set $u$ to be the vertex of smallest label in $G$, followed by $v_0$.
Then by the definition of $\mu(H)$ the vertex $u$ will be placed in $H_0$ which can be seen as forcing the gadget to false, satisfying the clause $y=0$ if the red edge $uy$ is selected, and satisfying the clause $x=1$ if the blue edge $ux$ is selected.
This concludes the proof.
\end{proof}

\section{Discussions}\label{sec:discussion}

In this work we have showed that enumerating all (resp.~minimal, maximal) signatures of an XOR-CNF formula is tractable, namely, Theorems~\ref{thm:xor-all-signatures}, \ref{thm:xor-max} and \ref{thm:2-xor-max}.
Observe that the algorithm of Theorem~\ref{thm:xor-max} runs in incremental-polynomial time while the other two run with polynomial delay.
Hence it is natural to ask whether Theorem~\ref{thm:xor-max} can be improved to polynomial delay. 

\begin{question}\label{qu:xor-delay}
Can the maximal signatures of an XOR-CNF be listed with polynomial delay?
\end{question}

The algorithm by Boros et al.~\cite{boros2003algorithms} underlying Theorem~\ref{thm:xor-max} relies on the enumeration of the circuits of a matroid, for which the same question seems open since two decades. 
Hence, answering Question~\ref{qu:xor-delay} would either require new techniques, or would need to answer the open question from \cite{boros2003algorithms} in the Boolean case.
A natural first step is to handle the case of $k$-XOR-CNF's for fixed values of $k\geq 3$.
Toward this direction, it is not clear whether the proof of Theorem~\ref{thm:2-xor-max} can be generalized, as it strongly relies on the underlying (edge-bicolored) graph defined by the formula.

Furthermore, except for the algorithm of Theorem~\ref{thm:xor-all-signatures}
that runs with polynomial space, the algorithms of Theorems \ref{thm:xor-max} and~\ref{thm:2-xor-max} require a space that is potentially exponential as solutions must be stored into a queue.
Consequently, a first natural question is the following.

\begin{question}\label{qu:2-xor-space}
    Can the maximal signatures of a 2-XOR-CNF be generated with polynomial delay and polynomial space?
\end{question}

We however stress the fact that a positive answer to Question~\ref{qu:2-xor-space} would improve the algorithm by Conte et al.~for maximal bipartite subgraphs enumeration for which the questions of achieving polynomial delay and space is open~\cite{conte2022proximity}.

Toward this direction, we now argue that flashlight search may not be adapted in order to give a positive answer to Question~\ref{qu:2-xor-space}, as the extension problem is hard in that case. 
Moreover, this result even holds when restricted to 2-XOR-CNF's consisting of disequalities, that is, in the context of maximal bipartite subgraph enumeration, which may be of independent interest.

\begin{theorem}\label{thm:extension}
The problem of deciding, given a graph $G$ and two disjoint edge sets $A,B\subseteq E(G)$, whether there exists a maximal bipartite (edge) subgraph $H$ of $G$ such that $A\subseteq E(H)$ and $B\cap E(H)=\emptyset$ is $\NP$-complete.
\end{theorem}

\begin{longproof}
We first argue that the problem belongs to \NP{}.
Note that bipartiteness is hereditary, meaning that every subgraph of a bipartite graph is bipartite.
Moreover since bipartiteness can be decided in polynomial time, we can check maximality of $H$ by trying to add every edge in $E(G)\setminus E(H)$ and checking whether the resulting graph is bipartite.
Now since the two extra conditions $A\subseteq E(H)$ and $B\cap E(H)=\emptyset$ can be checked in polynomial time, we conclude that $H$ itself is a polynomial certificate.

We now prove \NP-hardness by reduction from 3-SAT.
Let $\varphi=C_1\land \dots \land C_m$ be an instance of 3-SAT on variable set $V(\varphi)=\{v_1, \dots, v_n\}$.
We construct $G$ by creating two vertices $x_i,y_i$ representing the literals of variable $v_i$ for every $1\leq i\leq n$, a special vertex $u$, and a vertex $c_j$ representing the clause $C_j$ for every $1\leq j\leq m$.
We add the edges $x_iy_i$ for every $1\leq i\leq n$, the edge $x_ic_j$ if $v_i$ appears positively in $C_j$, and the edge $y_ic_j$ if it appears negated.
Finally, we connect $u$ to every other vertex in $G$ making it a universal vertex.
This completes the construction of the graph.
Note that it is not bipartite as it contains triangles $x_iy_iu$ for $1\leq i\leq n$.
The two sets $A,B\subseteq E(G)$ are defined as $A\coloneqq \{x_iy_i : 1\leq i\leq n\}$ and $B\coloneqq \{uc_j : 1\leq j\leq m\}$.
We illustrate the reduction in Figure \ref{fig:extension}.

\begin{figure}[ht!]
    \centering
    \includegraphics[scale=\FIGextension]{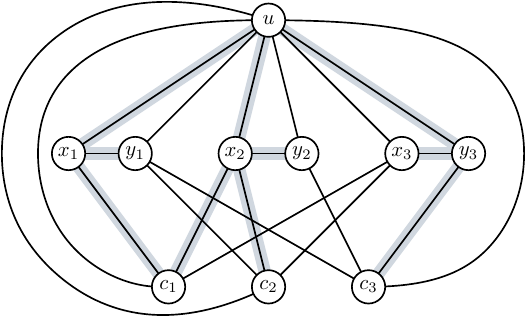}
    \caption{The reduction of Theorem \ref{thm:extension} with the CNF $\varphi = (v_1 \lor v_2 \lor v_3) \land (\bar{v_1} \lor v_2 \lor v_3) \land (\bar{v_1} \lor v_2 \lor \bar{v_3})$. We have $A = \{x_1 y_1, x_2 y_2, x_3 y_3\}$ and $B = \{c_1 u, c_2 u, c_3 u\}$. A maximal bipartite subgraph including $A$ and avoiding $B$ is highlighted in grey. It corresponds to the model of $\varphi$ which assigns $1$ to $v_1$ and $v_2$, and $0$ to $v_3$.}
    \label{fig:extension}
\end{figure}

Let us first show that there exists a maximal bipartite subgraph of $G$ containing $A$ and avoiding $B$ whenever $\varphi$ is satisfiable.
Consider a model $\a$ of $\varphi$ and consider the subgraph $H$ of $G$ defined as 
\begin{align*}
    E(H)\coloneqq  A 
    &\cup \{ux_i : \a(v_i)=1\} \cup \{uy_i : \a(v_i)=0\}\\
    & \cup \{c_jx_i : \a(v_i)=1\ \text{and}\ v_i\in C_j\}\\ 
    &\cup \{c_jy_i\hspace{.03cm} : \a(v_i)=0\ \text{and}\ \overline{v}_i\in C_j\}.
\end{align*}
Clearly $H$ extends subset $A$.
We first argue that it is bipartite.
Consider the bipartition $(V_1,V_2)$ where $V_1=\{x_i : \a(v_i)=1\}\cup \{y_i : \a(v_i)=0\}$ and $V_2=V\setminus V_1$.
By construction $V_1$ is edgeless.
The set $V_2$ is edgeless too as by the definition of $H$ vertices adjacent to $u$ or to $c_j$, $1\leq j\leq m$ in $H$ are precisely those in $V_1$.
Thus $H$ is bipartite.
We greedily extend $H$ into a maximal bipartite subgraph and note $H'$ the result.
Suppose toward a contradiction that $H'$ does not satisfy the required properties, i.e., that it contains an edge $uc_j$ for some $1\leq j\leq m$.
Then by definition of $V_1$ and the fact that $\a$ is a model of $\varphi$, we get that $uc_jw$ would induce a triangle for $w$ the neighbor of $c_j$ in $V_1$.
This contradicts the fact that $H'$ is bipartite.
Consequently $H'$ extends $A$ and avoids $B$ as required.

Suppose now that there exists a maximal bipartite subgraph $H$ of $G$ with $A\subseteq E(H)$ and $B\cap E(H)=\emptyset$.
Let $(V_1,V_2)$ be a bipartition of $H$.
Note that since $x_1y_1,\dots, x_ny_n$ belong to $A$ (hence to $H$), the sets $V_1$ and $V_2$ contain exactly one element from $\{x_i,y_i\}$ for each $1\leq i\leq n$.
Let us assume without loss of generality that $u\in V_2$.
Then since every edge $uc_j$, $1\leq j\leq m$ belongs to $B$, it must be that every such $c_j$ lies in~$V_2$ as well.
Thus, except for the choice of which of $\{x_i,y_i\}$ belongs to $V_1$, $1\leq i\leq n$, the bipartition $(V_1,V_2)$ is now completely characterized.
Consider a triangle $ux_iy_i$, $1\leq i\leq n$.
Then one of $x_i,y_i$ belongs to $V_1$, call it $\ell_1$, and the other belongs to $V_2$, call it $\ell_2$.
Clearly $u\ell_2$ is not an edge of $H$ as both $u$ and $\ell_2$ lie in the same side of the bipartition.
By assumption, $u\in V_2$ and $\ell_1\in V_1$ and the edge $u\ell_1$ is in $G$.
Since $H$ is maximal, this implies that $u\ell_1$ is an edge of $H$.
Thus the set of edges $u\ell$ in $H$ for $\ell\in V_1$ defines an assignment $\a$ of $\varphi$, where $\a(v_i)=1$ if $\ell=x_i$, and $\a(v_i)=0$ if $\ell=y_i$.
Hence the edges of $H$ having some $c_j$ for endpoint are precisely those connecting $c_j$ to $V_1$.
We claim that every such $c_j$ has at least one such incident edge, hence that $\a$ is a satisfying truth assignement of $\varphi$.
Indeed, if it was not the case then $c_j$ would be disconnected from the rest of the graph.
But in that case, we could have added $uc_j$ to the graph still maintaining the bipartiteness (changing $c_j$ to the other side of the bipartition) contradicting the maximality of $H$.
This concludes the proof.
\end{longproof}

Another question related to Question~\ref{qu:2-xor-space} concerns the existence of an output-poly\-nomial time and polynomial space algorithm for maximal signatures enumeration on general XOR-CNF's. Answering this would constitute a first step toward the same question for circuits enumeration in matroids.

Concerning prospective research directions, let us restate here the implicit question from~\cite{berczi2021generating} that motivated this work, which additionally was posed at the WEPA'19 open problem session, and that is still open to date.
Recall that a family of CNF's is tractable if there is a polynomial-time algorithm to decide the satisfiability of any formula (and its subformulas) in the family, even if some  of the variables are fixed to arbitrary values.

\begin{question}
Can the maximal signatures of tractable CNF's be enumerated in total-polynomial time?
\end{question}

Natural examples of tractable CNF's include 2-CNF's and Horn CNF's for which, to the best of our knowledge, no progress has been made.
Toward this direction, the case of Horn 2-CNF also seems open.

Another question left open by the work of Bérczi et al.~deals with the \emph{dimension} of the formula, defined as the maximum size of a clause it contains. 
Namely, the algorithm of \cite[Theorem 4]{berczi2021generating} only performs in incremental-polynomial time for fixed dimension, and it is open whether the same result can be obtained for formulas of arbitrary dimension.

Finally, another promising direction is the parameterized study of signatures enumeration.
More particularly, it can be seen that the algorithm of \cite[Theorem~4]{berczi2021generating} is \XP{}-incremental parameterized by the dimension $k$, that is, it generates the $i^\text{th}$ solution in time $(\|\varphi\|+i)^{f(k)}$ for some computable function $f$.
It would be interesting to know whether this algorithm can be improved to run with \FPT{}-delay, that is, to run with $f(k)\cdot \|\varphi\|^{O(1)}$ delay; see e.g.~\cite{elbassioni2008fpt,creignou2017paradigms,creignou2019parameterised,golovach2022refined,bartier2024hypergraph} for more details on the parameterized aspects of enumeration problems.
A preliminary step in that direction would be to get such running times for a double parameterization.
Among parameters, the maximum occurrence $\omega$ of a variable in the formula is natural.
This parameter is relevant as in \cite[Theorem~3]{berczi2021generating} the authors give a simpler algorithm for signatures enumeration when both $k$ and $\omega$ are bounded. 
Improving this \XP{}-incremental algorithm into one that run with \FPT{}-delay parameterized by $k$ plus $\omega$ seems open.

\paragraph{Acknowledgements.}  The authors would like to thank Yasuaki Kobayashi, Kazuhiro Kurita, Kazuhisa Makino and Kunihiro Wasa for preliminary discussions on the topic of this paper, as well as anonymous reviewers for valuable feedback on earlier versions of this manuscript.

\bibliographystyle{alpha}
\bibliography{main}

\end{document}